\documentclass[twocolumn]{aastex63}

\usepackage{amsmath}    
\usepackage{graphicx}   
\usepackage{verbatim}   
\usepackage{color}      
\usepackage{subfigure}  
 
\def\lowsim{\mathrel{\lower 0.7 ex \hbox to 0 pt{$\sim$\hss}}}
\def\lesssim{\lowsim<}

 \newcommand{\go}{{\gamma-1}}

 \newcommand{\dt}{\Delta t}
 
 \newcommand{\zp}{$z^+$}
 \newcommand{\zm}{$z^-$}

 \newcommand{\m}[1]{\mathbf #1} 

 \newcommand{\newadd}[1]{\textcolor{blue}{\bf #1}}

\begin{document}

\title{Measures of Scale Dependent Alfv\'enicity in the First PSP Solar Encounter}
\author[0000-0003-0602-8381]{T.~N. Parashar}
\email{tulasinandan@gmail.com}
\affiliation{Department of Physics and Astronomy, Bartol Research Institute, University of Delaware, Newark, DE 19716, USA}

\author[0000-0002-5317-988X]{M. L. Goldstein}
\affiliation{NASA Goddard Space Flight Center, Greenbelt, MD 20771, USA}
\affiliation{University of Maryland Baltimore County, Baltimore, MD 21250, USA}

\author[0000-0002-2229-5618]{B.~A. Maruca}
\affiliation{Department of Physics and Astronomy, Bartol Research Institute, University of Delaware, Newark, DE 19716, USA}

\author[0000-0001-7224-6024]{W.~H. Matthaeus}
\email{whm@udel.edu}
\affiliation{Department of Physics and Astronomy, Bartol Research Institute, University of Delaware, Newark, DE 19716, USA}

\author[0000-0003-3414-9666]{D. Ruffolo}
\affiliation{Department of Physics, Faculty of Science, Mahidol University, Bangkok 10400, Thailand}

\author[0000-0002-6962-0959]{R. Bandyopadhyay}
\affiliation{Department of Physics and Astronomy, Bartol Research Institute, University of Delaware, Newark, DE 19716, USA}

\author[0000-0002-7174-6948]{R. Chhiber}
\affiliation{NASA Goddard Space Flight Center, Greenbelt, MD 20771, USA}
\affiliation{Department of Physics and Astronomy, Bartol Research Institute, University of Delaware, Newark, DE 19716, USA}

\author[0000-0001-8478-5797]{A. Chasapis}
\affiliation{Laboratory for Atmospheric and Space Physics, University of Colorado Boulder, Boulder, CO 80303, USA}

\author[0000-0001-8358-0482]{R. Qudsi}
\affiliation{Department of Physics and Astronomy, Bartol Research Institute, University of Delaware, Newark, DE 19716, USA}

\author{D. Vech}
\affiliation{Climate and Space Sciences and Engineering, University of Michigan, Ann Arbor, MI 48109, USA}
\affiliation{Laboratory for Atmospheric and Space Physics, University of Colorado Boulder, Boulder, CO 80303, USA}

\author[0000-0001-6565-2921]{D. A. Roberts}
\affiliation{NASA Goddard Space Flight Center, Greenbelt, MD, USA}

\author[0000-0002-1989-3596]{S.~D. Bale}
\affiliation{Space Sciences Laboratory, University of California, Berkeley, CA 94720-7450, USA}
\affiliation{Physics Department, University of California, Berkeley, CA 94720-7300, USA}
\affiliation{The Blackett Laboratory, Imperial College London, London, SW7 2AZ, UK}

\author[0000-0002-0675-7907]{J.~W. Bonnell}
\affiliation{Space Sciences Laboratory, University of California, Berkeley, CA 94720-7450, USA}

\author[0000-0002-4401-0943]{T. Dudok de Wit}
\affiliation{LPC2E, CNRS and University of Orl\'eans, Orl\'eans, France}

\author[0000-0003-0420-3633]{K. Goetz}
\affiliation{School of Physics and Astronomy, University of Minnesota, Minneapolis, MN 55455, USA}

\author[0000-0002-6938-0166]{P.~R. Harvey}
\affiliation{Space Sciences Laboratory, University of California, Berkeley, CA 94720-7450, USA}

\author[0000-0003-3112-4201]{R.~J. MacDowall}
\affiliation{Code 695, NASA Goddard Space Flight Center, Greenbelt, MD 20771, USA}

\author[0000-0003-1191-1558]{D. Malaspina}
\affiliation{Laboratory for Atmospheric and Space Physics, University of Colorado Boulder, Boulder, CO 80303, USA}

\author[0000-0002-1573-7457]{M. Pulupa}
\affiliation{Space Sciences Laboratory, University of California, Berkeley, CA 94720-7450, USA}

\author[0000-0002-7077-930X]{J.~C. Kasper}
\affiliation{Climate and Space Sciences and Engineering, University of Michigan, Ann Arbor, MI 48109, USA}
\affiliation{Smithsonian Astrophysical Observatory, Cambridge, MA 02138 USA}

\author[0000-0001-6095-2490]{K.~E. Korreck}
\affiliation{Smithsonian Astrophysical Observatory, Cambridge, MA 02138 USA}

\author[0000-0002-3520-4041]{A.~W. Case}
\affiliation{Smithsonian Astrophysical Observatory, Cambridge, MA 02138 USA}

\author{M. Stevens}
\affiliation{Smithsonian Astrophysical Observatory, Cambridge, MA 02138 USA}

\author[0000-0002-7287-5098]{P. Whittlesey}
\affiliation{Space Sciences Laboratory, University of California, Berkeley, CA 94720-7450, USA}

\author{D. Larson}
\affiliation{Space Sciences Laboratory, University of California, Berkeley, CA 94720-7450, USA}

\author{R. Livi}
\affiliation{Space Sciences Laboratory, University of California, Berkeley, CA 94720-7450, USA}

\author[0000-0002-2381-3106]{M. Velli}
\affiliation{Department of Earth, Planetary, and Space Sciences, University of California, Los Angeles, CA 90095, USA}

\author[0000-0003-2409-3742]{N. Raouafi}
\affiliation{Johns Hopkins University Applied Physics Laboratory, Laurel, MD, USA}

\begin{abstract}
The solar wind shows periods of highly Alfv\'enic activity, where velocity fluctuations and magnetic fluctuations are aligned or anti-aligned with each other. It is generally agreed that solar wind plasma velocity and magnetic field fluctuations observed by Parker Solar Probe (PSP) during the first encounter are mostly highly Alfv\'enic. However, quantitative measures of Alfv\'enicity are needed to understand how the characterization of these fluctuations compares with standard measures from prior missions in the inner and outer heliosphere, in fast wind and slow wind, and at high and low latitudes. To investigate this issue, we employ several measures to quantify the extent of Alfv\'enicity -- the Alfv\'en ratio $r_A$, {normalized} cross helicity $\sigma_c$, {normalized} residual energy $\sigma_r$, and the cosine of angle between velocity and magnetic fluctuations $\cos\theta_{vb}$. We show that despite the overall impression that the Alfv\'enicity is large in the solar wind sampled by PSP during the first encounter, during some intervals the cross helicity starts decreasing at very large scales. These length-scales (often $> 1000 d_i$) are well inside inertial range, and therefore, the suppression of cross helicity at these scales cannot be attributed to kinetic physics. This drop at large scales could potentially be explained by large-scale shears present in the inner heliosphere sampled by PSP. In some cases, despite the cross helicity being constant down to the noise floor, the residual energy decreases with scale in the inertial range. These results suggest that it is important to consider all these measures to quantify Alfv\'enicity.
\end{abstract}

\section{Introduction}
The low frequency, magnetofluid-scale turbulence observed in the solar wind is often described as ``Alfv\'enic'', referring to the often-seen high degree of correlation between velocity and magnetic field fluctuations \citep{BelcherJGR71}. This significant Alfv\'enic correlation is often attributed more to high latitude wind \citep{McComasJGR00} or to high speed low-latitude wind \citep{BrunoEA03}, and generally more to distances closer to the Sun rather than farther. However, there are many exceptions, and high Alfv\'enicity intervals can sometimes be observed in slow low-latitude intervals, or at large heliocentric distances \citep{RobertsEA87b}. Nevertheless the prevailing expectation for Parker Solar Probe (PSP), as it approached closer to the Sun than any previous spacecraft, was almost certainly that it would observe highly Alfv\'enic fluctuations. Indeed, most reports of the first two encounters (this volume) at least qualitatively describe the fluctuations, even the ``jets'' or ``switchbacks,'' as having an Alfv\'enic character \citep{Bale:prep}. Here we will probe more deeply into the nature of the Alfv\'enic correlation in the first solar encounter of PSP \citep{FoxSSR16}, examining several independent measures of Alfv\'enicity, and resolving the associated correlations according to length scales. Recognizing that the first encounter may not be entirely typical \citep{Kasper2019a:prep}, we will argue that the departures from pure Alfv\'enicity recorded in the inner heliosphere by PSP may provide clues as to the dynamics at work in this turbulent plasma so close to the corona. 
 
Alfv\'enicity is an important concept in plasma dynamics, but the precise meaning of this terminology is ambiguous without some clarification. In fact, it has been used to refer to different (although related) constructs by different authors. A first major issue is the existence of different quantitative measures of the ``Alfv\'enic property'' \citep{BelcherJGR71}. As commonly defined, these are the Alfv\'en ratio $r_A$, the cross helicity $\sigma_c$, the residual energy $\sigma_r$, and the angle of alignment between velocity and magnetic field fluctuations $\cos\theta$. Each of these measures is associated with Alfv\'enicity and may further be defined locally, or by regional averages, or scale (filtered) averages, or a global/ensemble average. For purposes of definition we employ $\langle \dots \rangle$ to denote an ensemble average.

The cross helicity $H_c = \langle {\bf v }\cdot {\bf b}\rangle$, where ${\bf v}, {\bf b}$ are velocity and magnetic field fluctuations, is a rugged invariant of ideal incompressible magnetohydrodynamics (MHD). The physical significance of $H_c$ is revealed by comparing it with another ideal invariant, the incompressible fluctuation energy density per unit mass, $E = E_b+ E_v = \langle |{\bf v}|^2 \rangle/2 +  \langle |{\bf b}|^2 \rangle/2$. The dimensionless measure is {the normalized cross helicity} $\sigma_c = H_c/E$ such that $-1 \le \sigma_c \le 1$. Fluctuations with large $|\sigma_c| \to 1$ are sometimes described as being Alfv\'enic. Note that for convenience, the magnetic fluctuation  ${\bf b}$  is usually measured in Alfv\'en speed units, i.e., with implied division by $\sqrt{\mu_0 n_p m_p}$. An important property is that, by definition, Alfv\'en waves have ${\bf v} = \pm {\bf b}$ 
and therefore such waves have $\sigma_c = \pm 1$ by definition. One may also note that, in terms of the Els\"asser variables ${\bf z} ^\pm = {\bf v} \pm {\bf b}$, the normalized cross helicity may be written in the revealing form $\sigma_c = \langle |{\bf z}^{+}|^2 - |{\bf z}^-|^2 \rangle / \langle |{\bf z}^{+}|^2 + |{\bf z}^-|^2 \rangle$.

The ``Alfv\'en ratio'' is the ratio of flow kinetic energy to magnetic fluctuation energy, $r_A = \langle |{\bf u}|^2 \rangle /\langle |{\bf b}|^2 \rangle$. Its physical significance is to measure the degree of energy equipartition of flow and magnetic fluctuations. A single Alfv\'en wave has $r_A=1$, and a random phase mixture of small or large amplitude Alfv\'en waves will exhibit equipartition with $r_A = 1$. For this reason turbulence with energy equipartitioned in this sense is sometimes described as Alfv\'enic turbulence. Another related measure to quantify the relative energy in kinetic and magnetic fluctuations is the {normalized} residual energy $\sigma_r = (\langle |\bf{u}|^2 \rangle - \langle |\bf{b}|^2 \rangle)/(\langle |\bf{u}|^2 \rangle + \langle |\bf{b}|^2 \rangle)$. {In the rest of the paper, for brevity, we will drop the ``normalized'' prefix from cross helicity and residual energy, with the understanding that these imply the normalized versions $\sigma_c$ and $\sigma_r$.}

Finally the alignment cosine of the angle $\theta$ between the fluctuations in ${\bf v}$ and $\bf b$ may be written as $\cos \theta_{vb} = {\bf v} \cdot {\bf b}/(|{\bf v}| |{\bf b}|)$. The global alignment cosine is $\cos \Theta  \equiv \langle  {\bf v} \cdot {\bf b}\rangle / [\langle |{\bf v}|^2 \rangle \langle |{\bf b}|^2 \rangle]^{1/2} = H_c/2 \sqrt{E_vE_b}$. Note that $\cos \Theta$ is {\it not} a ratio of ideal global invariants. Nevertheless it is a quantity often discussed in connection with Alfv\'enicity, and turbulence with large values of $\cos{\Theta}$ is sometimes referred to as  Alfv\'enic turbulence.

The above measures of Alfv\'enicity are not independent. They are related by the well known identities, $\sigma_c = 2 \cos{\theta_{vb}} {\sqrt{r_A}}/{(1+r_A)}$, and $\cos \theta_{vb} = \sigma_c/\sqrt{1-\sigma_r^2}$. Thus, for example, perfectly directionally-aligned fluctuations are necessarily of pure cross helicity only if they are in energy equipartition. A complete picture of Alfv\'enicity of an interval requires addressing as many of these measures together as possible. 

Beyond these kinematic measures of Alfv\'enicity, there are at least three dynamical scenarios related to these physical properties: these are global dynamic alignment over time \citep{DobrowolnyAA80,MatthaeusANYAS80}, scale dependent dynamic alignment \citep{BoldyrevPRL06, MasonPRL06, BoldyrevApJL09}, and patchy alignment in real space \citep{MilanoPP01, MatthaeusPRL08}. All of these constructs have been studied in separate contexts over the last few decades. Each employs the measures $r_A$, $\sigma_c$, $\sigma_r$, and $\cos{\theta}$, or equivalent measures, in various averages and measures, to characterize Alfv\'enic correlation and Alfv\'enic turbulence

Before turning to new results, it is important to establish the observational context. Alfv\'enic fluctuations have typically been seen as a prominent feature of MHD-scale fluctuations in the inner heliosphere, for example in Mariner \citep{BelcherJGR71} and Helios \citep{BrunoJGR85, MarschJGR90} observations. Moving further outward, there is a general decrease in occurrence of very high cross helicity at low latitudes, although high Alfv\'enicity has been observed as far out as 9 au \citep{RobertsJGR87}. However at the higher latitudes explored by Ulysses \citep{BavassanoJGR98, BavassanoAIPC99, BreechJGR08} the Alfv\'enicity persists to out to larger distances than typically seen at lower latitudes. 
{A point of general consensus is that Alfvenicity decreases primarily due to shear \citep{RobertsJGR87, RobertsJGR92, zank1996evolution} with persistent contributions also due to expansion \citep{ZhouJGR90, OughtonJGR95}. More recent studies have further examined effects of shear on Alfvenicity employing more complete theoretical formulations \citep{BreechJGR08, ZankApJ12, AdhikariApJ15, ZankApJ17}.}

\begin{figure}
   \includegraphics[width=0.94\columnwidth]{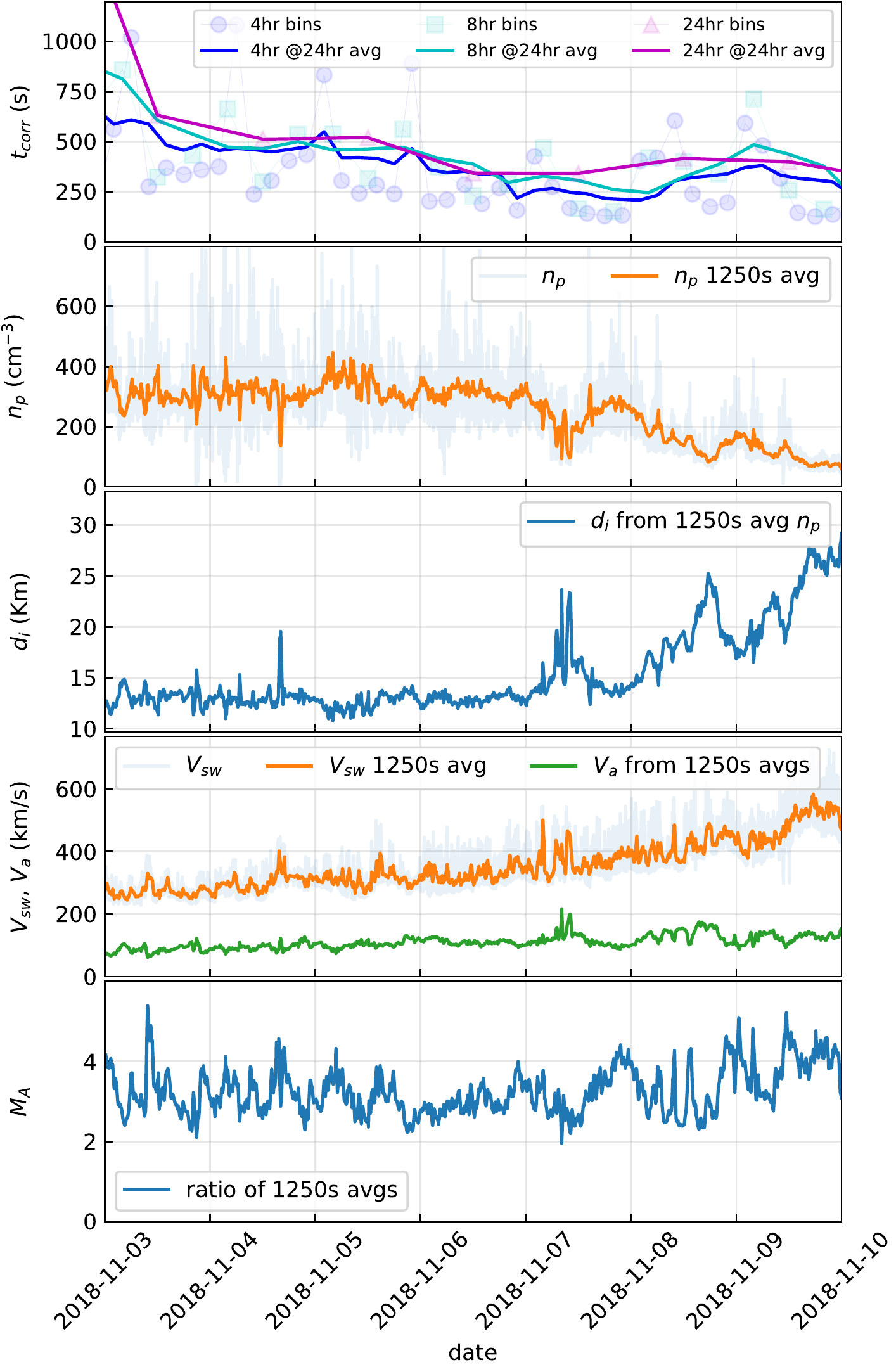}
   \caption{Overview of some key quantities of solar wind fluctuations as a function of time during the first PSP solar encounter: magnetic field correlation time $t_{corr}$, proton density $n_p$, ion inertial length $d_i$, and solar wind speed $V_{sw}$ and Alfv\'en speed $V_a$ and their ratio.}
   \label{1250_rolling}
\end{figure}

As far as spatial distribution is concerned, there have been a number of reports \citep{MilanoPP01, MatthaeusPRL08} that cross helicity tends to be found in organized patches, an effect apparently related to local turbulent relaxation. This effect is also consistent with 
solar wind observations \citep{OsmanApJ11}. A related concept is the scale dependence of cross helicity at MHD scales \citep{BoldyrevPRL06}. One interesting effect is related to the disparity of time scales in high cross helicity states: when $z^+ \gg z^-$ the time scale for transfer of the ``majority species'' $z^+$ becomes large compared to the time  scale for advection of the minority species $z^-$. Consequently the initial transfer from a large scale Alfv\'enic spectrum to small scales tends to be dominated by the {\it weaker} Els\"asser energy \citep{MatthaeusPRL83}. When present, this effect accelerates the overall amplification of dimensionless Alfv\'enicity $\sigma_c$, which is frequently, but not always seen in simulations of turbulent relaxation \citep{StriblingPFB91}. The exceptional cases, when this {\it dynamic alignment} does {\it not} occur are often associated with turbulence in which a substantial amount of energy is found in {\it velocity shears.} The idea that shear destroys an initial spectrum of high cross helicity by injecting equal amounts of the two Els\"asser energies has been investigated in both simulations and observations \citep{GoldsteinSSPP89, RobertsJGR92}. In these studies shear reduced cross helicity that initially was at the scale of the initial shear and over time the effect then spread across all scales. For quasi-steady cases, the alignment measured by $\cos{\theta_{vb}}$ has been conjectured to increase with decreasing scales, leading to a modification of the cascade theory \citep{BoldyrevPRL06, MasonPRL06} on which alignment progressively increases with decreasing scale. 

One may also ask what happens to the three measures of Alfv\'enicity in a kinetic plasma environment. This has been recently studied using MMS data and kinetic PIC simulation \citep{ParasharPRL18}. For sample intervals that are Alfv\'enic in the inertial range, 
MMS data show that $\sigma_c$ starts at a non-zero value at inertial range scales and approaches zero at kinetic scales, indicating lack of alignment between $\m{v}$ and $\m{b}$ at kinetic scales. This result is confirmed by comparison of multi-spacecraft estimates and single-spacecraft estimates. Preliminary study of PSP data \citep{VechApJ19} has also examined cross helicity and related alignments  at higher frequencies (smaller scales) approaching the kinetic range. Similar results to those of \citep{ParasharPRL18} are found. These results show the diminishing importance of cross helicity and alignment at or near ion inertial scales, which is not entirely surprising since $H_c$ is not an ideal invariant for kinetic plasmas;
in fact, even in Hall-MHD, one must consider a generalized helicity, and not the standard MHD cross helicity \citep{TurnerIEEETPS86}

\begin{figure*}[!hbt]
   \includegraphics[width=\textwidth]{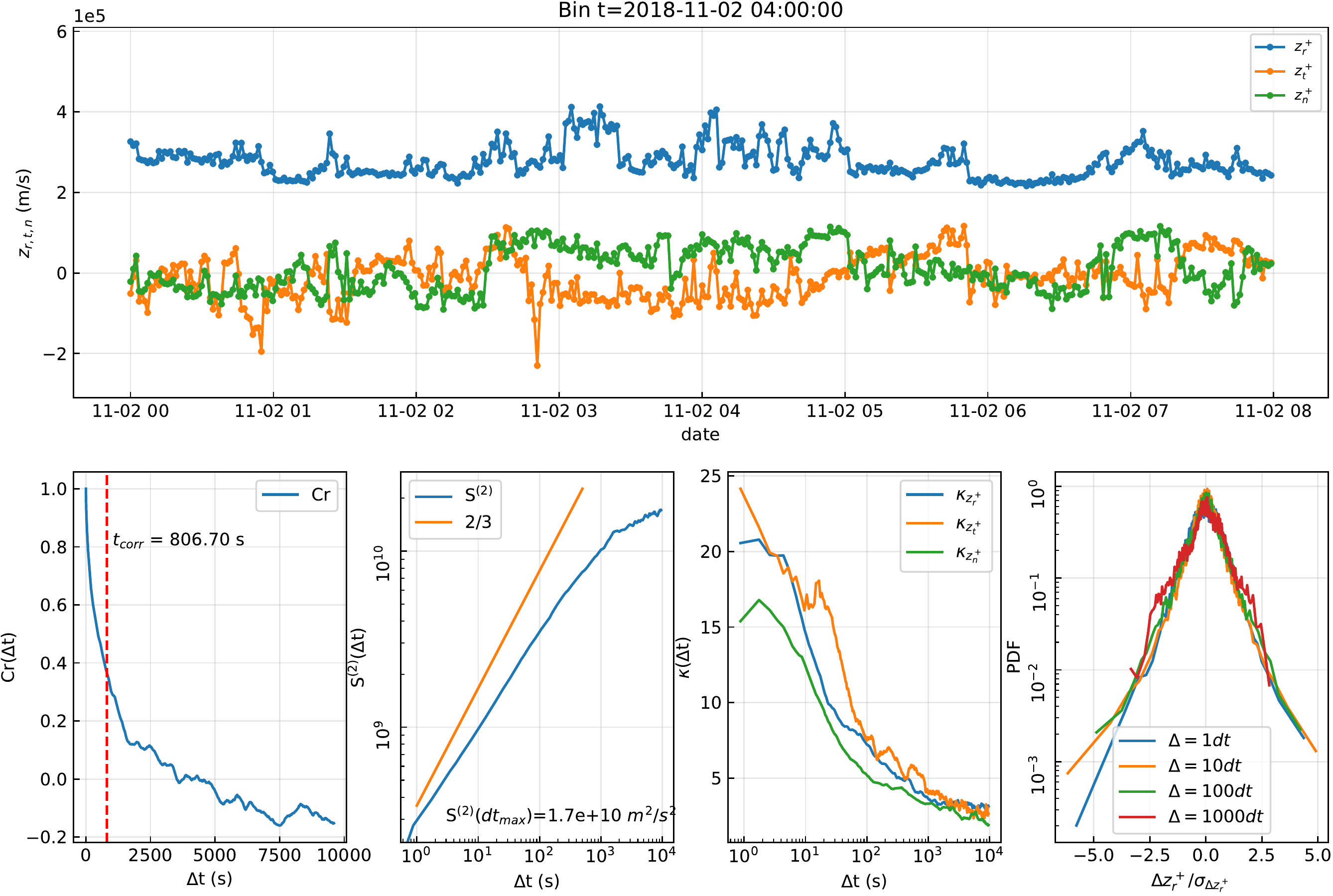}
   \caption{Overview of the turbulence properties of $z^+$ for an 8 hour period. Top panel shows time series of \zp components. Panels in the bottom row show: autocorrelation coefficient, second order structure function, scale dependent kurtosis, and PDFs of increments. See text for details.}
   \label{zp_overview}
\end{figure*}

Prior studies provide ample evidence for a variety of different possible scenarios involving cross helicity, ranging from local amplification, scale dependent increase through the inertial range, and decrease due to shear, expansion and kinetic effects. Nevertheless, there appears to be a general tendency to assume that Alfv\'enic fluctuations at MHD scales are more prevalent and more purely outward in the inner heliosphere. 
For this reason, much of the early discussion of MHD fluctuations in  the first PSP orbit has focused on relatively larger scale features that are Alfv\'enic. Here we examine this characterization in greater detail. In particular, in this study we are interested in behavior of $\sigma_c$, $\sigma_r$, $r_A$ and $\cos{\theta_{vb}}$ at relatively large inertial range scales in the inner heliosphere sampled by PSP. We show that in some cases, even when $\sigma_c$ remains constant through the inertial range, $\sigma_r$, and $r_A$ change significantly in the inertial range. In some other intervals, $\sigma_c$ decreases with scale in the inertial range ($> 1000 d_i$) in the inner heliosphere. This decline at scales much larger than ion kinetic scales rules out an explanation in terms of the kinetic physics as explored in \citet{ParasharPRL18}.

\section{Data \& Processing}
\begin{figure*}[!hbt]
   \includegraphics[width=\textwidth]{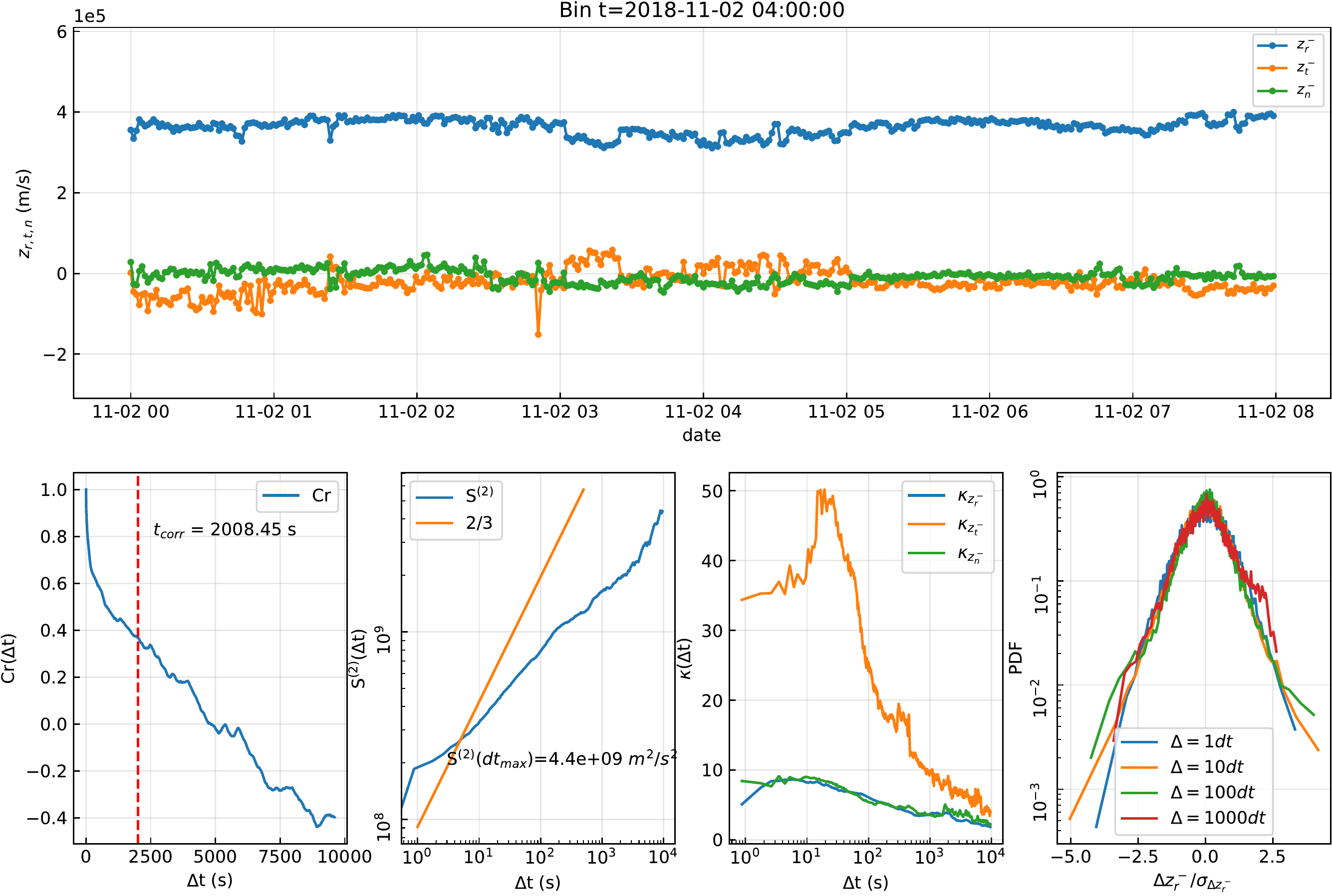}
   \caption{Overview of the turbulence properties of $z^-$ for an 8 hour period. Top panel shows time series of \zm\  components. Panels in the bottom row show: autocorrelation coefficient, second order structure function, scale dependent kurtosis, and PDFs of increments. The lack of energy in \zm\ fluctuations compared to \zp\ fluctuations is evident in suppressed fluctuations, and in the second order structure function.  See text for details.}
   \label{zm_overview}
\end{figure*}

PSP's first perihelion occurred on 2018/11/06 with high time-cadence data collection occurring between 2018/10/31-2018/11/11. The initial and final days did not have full coverage of high time-cadence data, so we choose to perform the analysis on data obtained between 2018/11/01 and 2018/11/10. Level-two PSP/FIELDS and Level-three data from the PSP/SWEAP archives are used for the analysis. Specifically, data are from the FIELDS flux-gate magnetometer (MAG) \citep{BaleSSR16} and Solar Probe Cup (SPC) \citep{KasperSSR16}. The time cadence of SPC varied during the encounter between 1 NYHz and 4 NYHz, where 1 NYHz is the inverse of 1 NYs (=0.874 s). To create a uniform time series, we resampled all data (SPC and fields) to 1 NYHz 
cadence. \newadd{Plasma data used are obtained by fitting Maxwellian distribution functions to SPC data.} Some unphysical spikes in SPC data, which are remnants of bad fits, are removed using a modified Hampel filter in the time domain \citep{BandyopadhyayEA18}. \newadd{The case studies presented in this analysis are from Nov 2 and Nov 4 in 2018, just before the first encounter.}

The resampled data are divided into subsets of various sizes (4 hr, 8 hr, and 24 hr) and the correlation time is computed for the magnetic field {as the time when the autocorrelation function is reduced by $1/e$}. The correlation time $\tau_{corr}$ is shown for each 4-, 8-, or 24-hr sub-interval as points in the top panel of Figure~\ref{1250_rolling}. The solid lines represent 24-hr running averages of these points. The correlation time typically depends on the averaging interval, and can be sensitive to larger scale fluctuations \citep{MatthaeusJGR82b, IsaacsJGR15, JagarlamudiApJ19}. The correlation times computed from intervals 4 hr or longer are all comparable to each other and fluctuate between 300s-600s. This number is consistent with the spectral break point between the $f^{-1}$ range and the inertial range \citep{ChenAPJ19}.

The computation of Els\"asser variables requires conversion of magnetic field fluctuations to Alfv\'enic velocity. This conversion is performed with some care. Large local variations of density do not imply a possibility of different point-wise Alfv\'en waves. An inertial range 
Alfv\'en wave and corresponding Alfv\'en speed should be defined over a reasonably large scale, one 
over which an MHD Alfv\'en wave can exist and propagate. Hence, we use density averaged over a few correlation times to convert magnetic field fluctuations into Alfv\'enic speeds. Here $\tau_{corr} \sim 300-600 s$ implies that a rolling average of 1250 s covers scales between 2 to 4 $\tau_{corr}$ over the encounter. The second panel of Figure \ref{1250_rolling} shows instantaneous density in light gray and the 1250s rolling average. This rolling average is used to define the Alfv\'en speed and the proton inertial scale $d_i$, and for conversion of magnetic fluctuations to velocity units. 

A comparison of the solar wind speed to Alfv\'en speed computed this way {gives an Alfv\'enic Mach number $M_A=V_{sw}/V_a \sim 3-4$}, marginally allowing us to use Taylor's hypothesis. A detailed study of Taylor's hypothesis for this encounter will be reported elsewhere.

\section{Results}
\begin{figure*}
   \includegraphics[width=\textwidth]{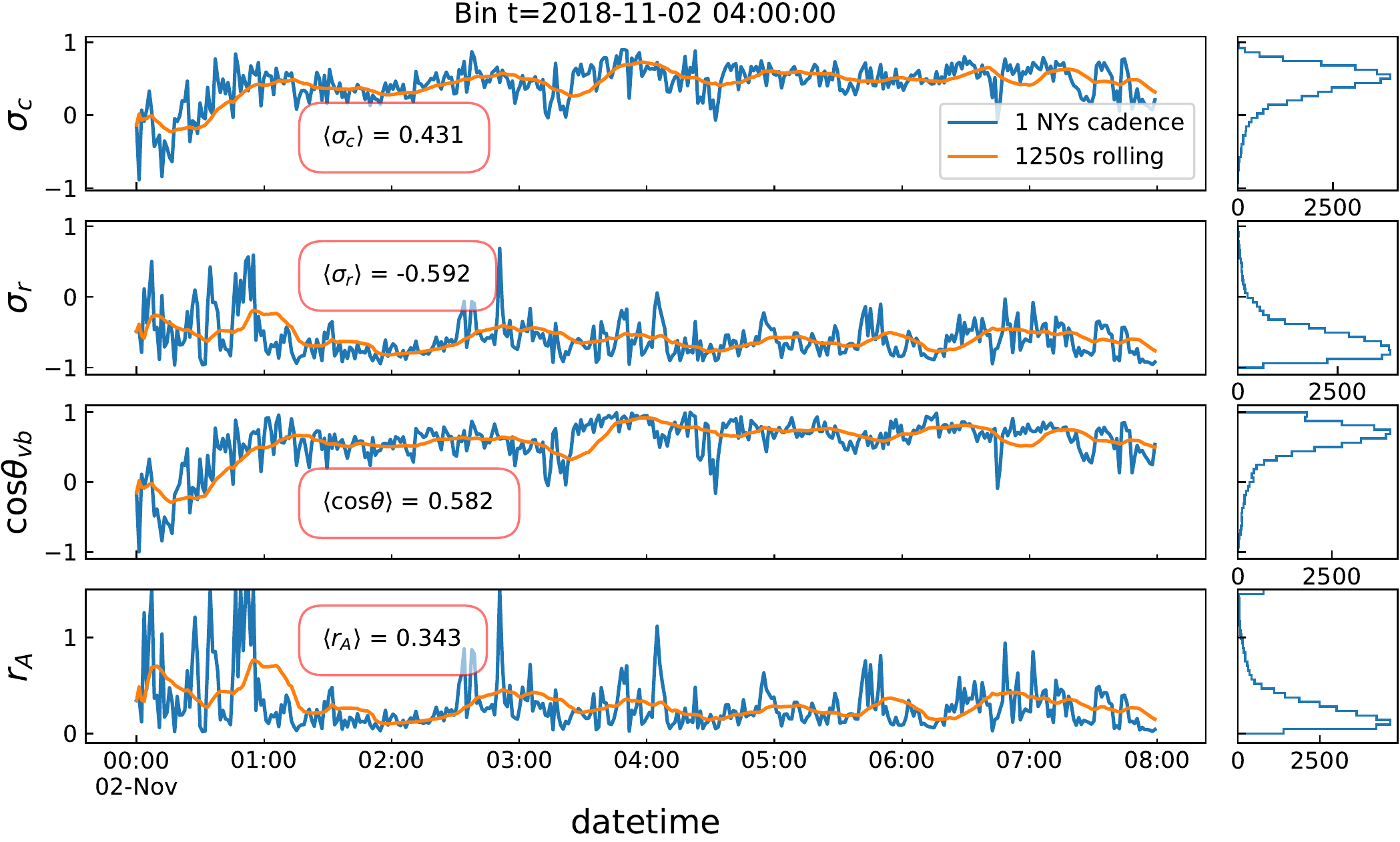}
   \caption{Various measures of Alfv\'enicity for an 8-hr window. Blue lines show the actual time series and orange lines show 1250-s running averages of the quantities. Histograms on the right share the $y$-axis with the left panels, and show the frequency of occurrence.
   The average values over the entire 8-hr data sample are shown in the boxes.  The average cross helicity (0.431), the average cosine (0.582), and the average Alfv\'en ratio (0.343) indicate a moderate or incomplete degree of Alfv\'enicity. By way of contrast, some Helios intervals have cross helicity above 0.95 \citep[see, e.g.,][]{RobertsEA87b, MarschPIH291, StansbyMNRAS18}.}
   \label{sc_overview}
\end{figure*}

Using the reprocessed data we compute Els\"asser variables and the relevant  quantifiers of Alfv\'enicity -- cross helicity $\sigma_c$, residual energy $\sigma_r$, Alfv\'en ratio $r_A$, and the alignment cosine $\cos \theta_{vb}$. Figures \ref{zp_overview} and \ref{zm_overview} show an overview of \zp\ and \zm\ fluctuations, respectively, for an 8-hr period centered at 2018/11/02-04:00:00. In each figure, the top panel shows the overview time series, and the four panels below it show the autocorrelation function $C(\dt) \equiv \langle | {\bf z}^\pm (t+\dt) \cdot {\bf z}^\pm(t)| \rangle/\langle {\bf z}^{\pm2} \rangle$, second-order structure function $D^{(2)}(\dt) \equiv \langle |\delta {\bf z}^\pm(t,\dt)|^2 \rangle$, scale-dependent kurtosis for individual components in the RTN coordinate system $\kappa_{r,t,n}(\dt) \equiv \langle |\delta z^\pm_{r,t,n}(t,\dt)|^4 \rangle/\langle |\delta z^\pm_{r,t,n}(t,\dt)|^2 \rangle^2$, and probability density functions (PDFs) of increments for four different increments of 1, 10, 100, and 1000 $dt$, where an increment is defined as $ \delta z_{r,t,n}^\pm(t,\dt) = z_{r,t,n}^\pm (t+\dt) - z_{r,t,n}^\pm(t)$, the time cadence is $dt$, 
and $\langle \ldots \rangle$ denotes averaging over $t$. 

In these Figures, \zp\ shows strong turbulent fluctuations, with a well-developed power spectrum as indicated by the second-order structure function. {Kolmogorov slope of $2/3$, typically observed in (magneto)hydrodynamic turbulence \citep{BiskampBook03}, is shown for reference. In this particular interval the slope is slightly different from the Kolmogorov value but in a significant number of intervals analyzed (not shown) the slope was close to $2/3$.} The correlation time for \zp\ is $\tau_{corr}\sim 800$ s, consistent with a roll over of the second order structure function at a few $\tau_{corr}$. The scale dependent kurtosis for \zp\ keeps increasing down to very small scales, while for \zm\ the peak of kurtosis occurs between 10-100s. {The decrease in kurtosis for \zm is likely because the signal is weaker for \zm and hence the noise becomes significant at larger scales.} The PDFs of increments show non-Gaussian features deep into the inertial range. The weaker Els\"asser field \zm,  on the other hand, shows suppressed turbulent fluctuations, a smaller correlation time, and about an order of magnitude smaller energy compared to \zp\ fluctuations. This behavior is consistent with outward-propagating Alfv\'enic fluctuations. We now discuss the individual measures of alignment both in time and as a function of scale. 

Figure \ref{sc_overview} shows various measures of Alfv\'enicity for an 8-hr interval during the encounter. Blue lines show the actual time series, orange lines show a 1250-s running average of the quantities. Histograms on the right show the frequency of occurrence of certain values. The average values of these quantities ($\langle\sigma_c\rangle \sim 0.43$, $\langle\sigma_r\rangle \sim-0.59$, $\langle\cos{\theta_{vb}}\rangle \sim 0.58$, and $\langle r_A\rangle\sim 0.34$) indicate a moderate or incomplete degree of alignment. Although most of the population has a fairly high $\sigma_c$, and $\cos\theta_{vb}$, locally the cross helicity shows large deviations from the mean value at time scales of the order of a few minutes.  This is consistent with locally patchy behavior of cross helicity as reported by \cite{MatthaeusPRL08} and \cite{OsmanApJ11} where it was shown that the cross helicity can show large systematic departures from the global average in localized patches. 


\begin{figure}[!hbt]
   \includegraphics[width=\columnwidth]{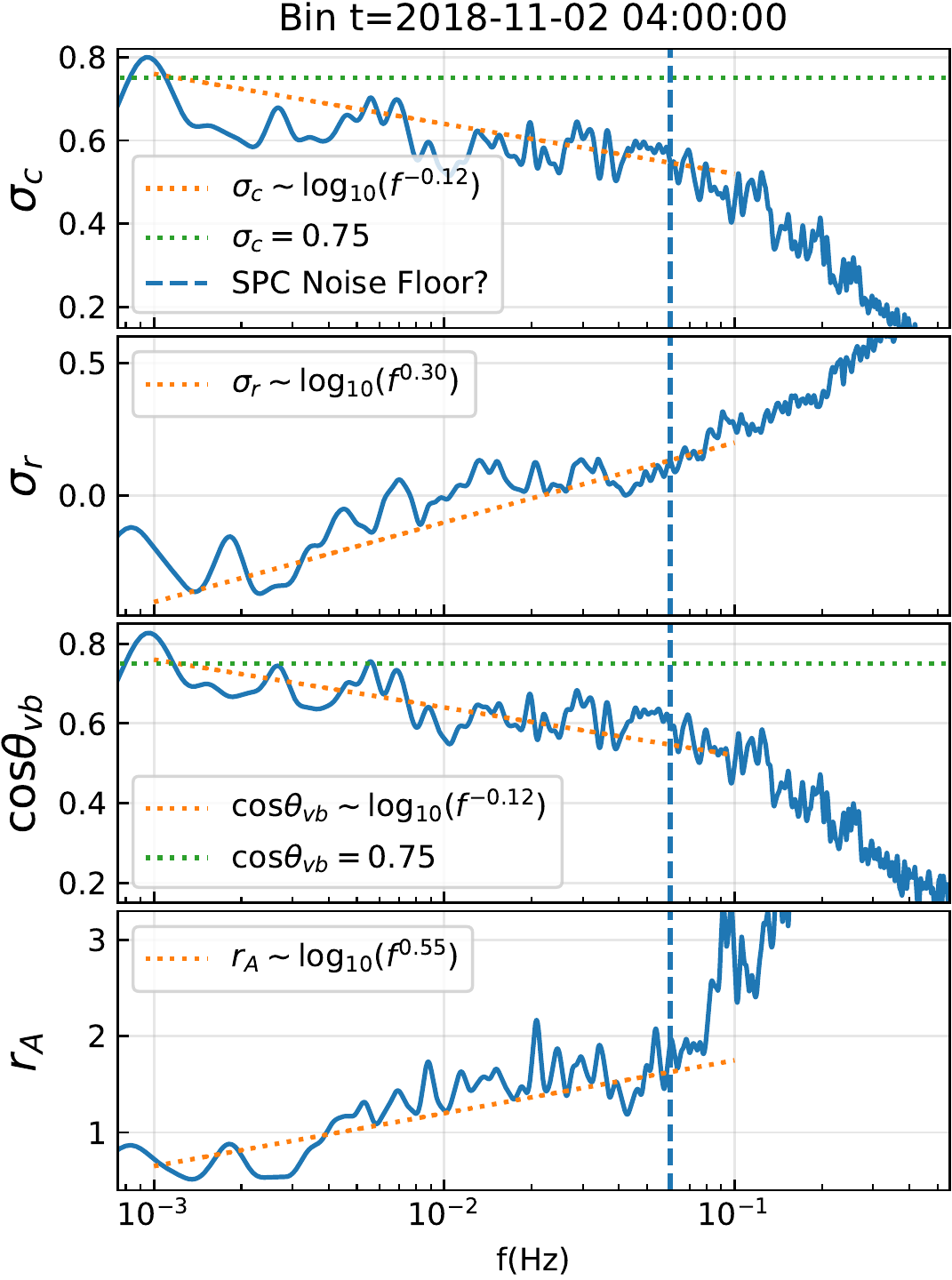}
   \caption{Various measures of Alfv\'enicity for the 8 hour window centered at 2018-11-02 04:00:00, as a function of scale. Vertical dashed line represents a frequency at which noise possibly becomes important, identified as the frequency where velocity spectrum starts flattening. The dotted green and orange lines in the top (cross helicity) panel show constant cross helicity and logarithmic decline respectively. It is evident that the cross helicity shows a logarithmic decline in the inertial range. Studies at 1 au show a steep decline in cross helicity close to kinetic scales \citep{ParasharPRL18, VerdiniApJ18}. However, the decline at large, MHD scales has also been observed in Helios data \citep{TuGRL90, BrunoAIPC96} and studied in the context of destruction by velocity shears \citep{GoldsteinSSPP89,RobertsJGR92}.}
   \label{sc_scale07}
\end{figure}

\begin{figure}[!hbt]
   \includegraphics[width=\columnwidth]{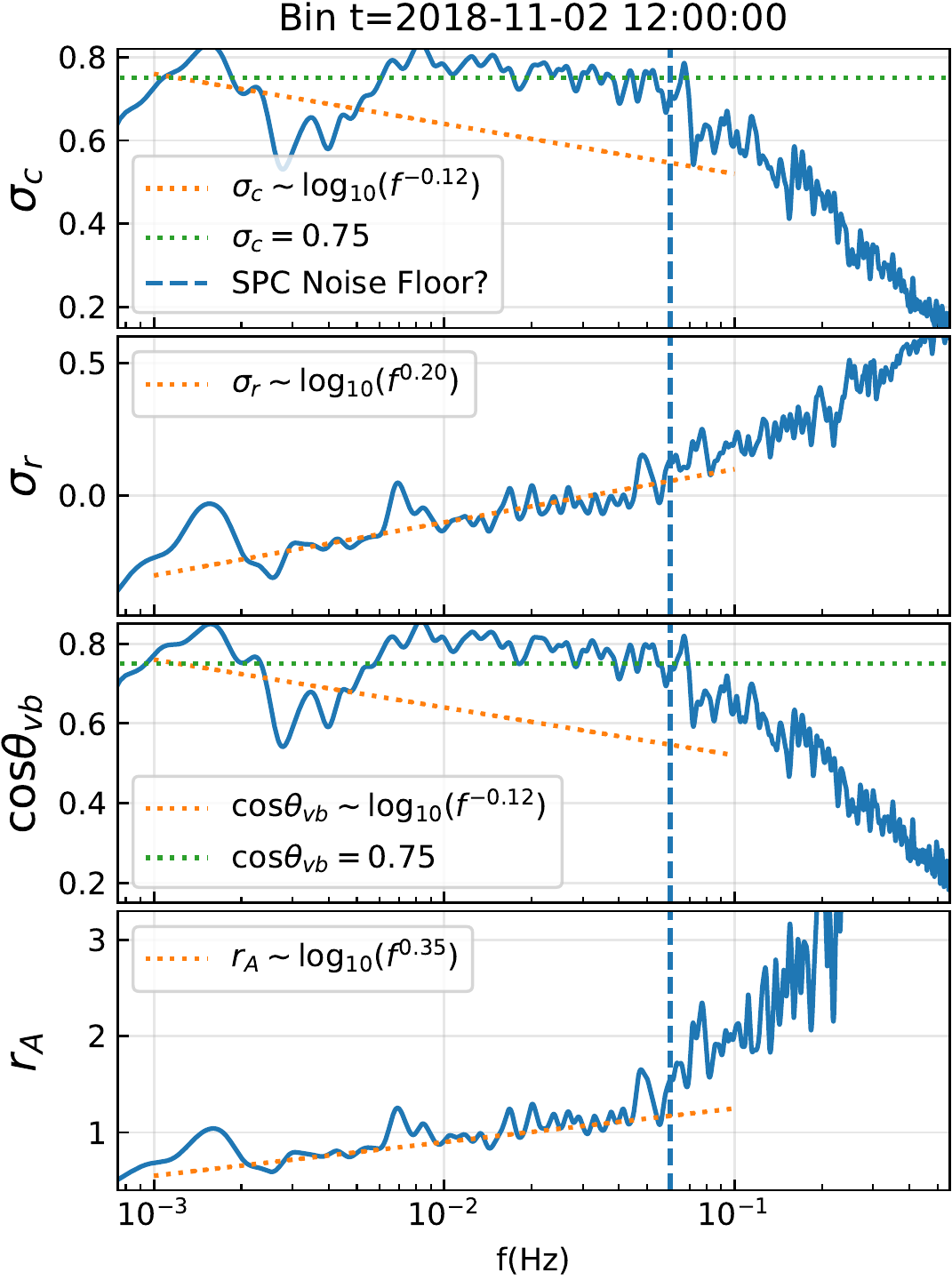}
   \caption{Various measures of Alfv\'enicity for an 8 hour bin centered at 2018-11-02 12:00:00, as a function of scale. Vertical dashed line represents a frequency at which noise possibly becomes important, identified as the frequency where velocity spectrum starts flattening. The dotted green and orange lines in the top (cross helicity) panel show constant cross helicity and logarithmic decline respectively. It is evident that the cross helicity in this interval is constant down to the noise floor. However, the Alfv\'en ratio and the residual energy both show monotonic {reduction in magnetic dominance, as evidenced by the overplotted logarithmic trends.}}
   \label{sc_scale08}
\end{figure}

\begin{figure}[!hbt]
   \includegraphics[width=\columnwidth]{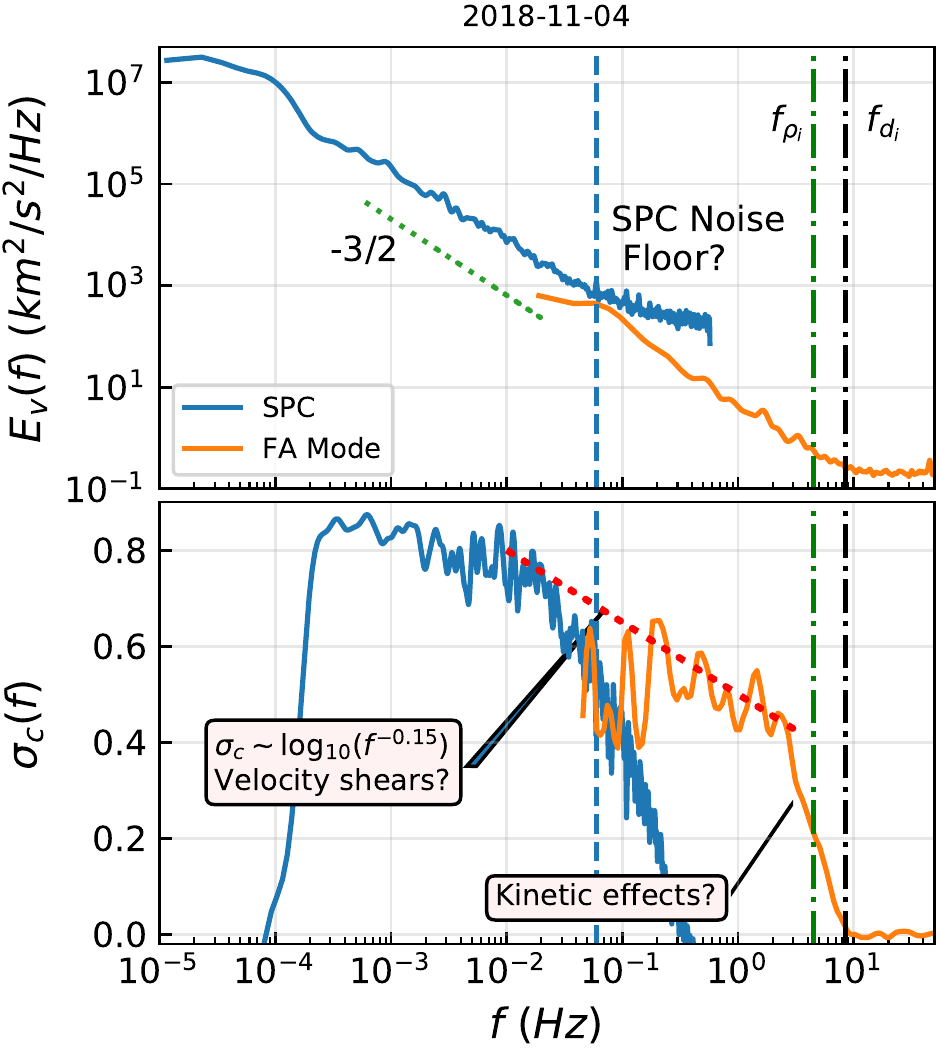}
   \caption{Extension of spectral range using combined normal and FA modes of SPC. Top panel shows the velocity spectra from the two modes using data for the full day on November 4, 2018. {Vertical dot-dashed lines represent $f_{\rho_i} = V_{sw}/(2\pi\rho_i)$ and $f_{d_i} = V_{sw}/(2\pi d_i)$, where $\rho_i$ and $d_i$ are the gyroradius and inertial length of a proton.} Combined, these instruments cover almost five decades in spectral range (excluding the largest decade that is affected by the windowing function). In the same range, cross helicity shows a logarithmic decline starting at a few hundred $d_i$. Close to kinetic scales the decline is very sharp, consistent with observations at 1 au.}
   \label{parashar_vech}
\end{figure}

To get a more complete picture of Alfv\'enicity, we perform a scale decomposition of these alignment measures. Figure \ref{sc_scale07} shows the Fourier spectra of $\sigma_c$, $\sigma_r$, $\cos \theta_{vb}$, and $r_A$ as a function of frequency. The vertical dashed line marks the frequency where noise becomes important, identified by flattening of velocity spectra (not shown). All measures of Alfv\'enicity show departures from large scale values in the inertial range. The decline is approximately logarithmic, as suggested by the orange dashed line in the top panel for $\sigma_c$. Similar logarithmic changes are seen for $\cos \theta_{vb}$, $\sigma_r$, and $r_A$. \newadd{The apparent discrepancy in the scales where $\sigma_r$ crosses zero and $r_A$ crosses one is purely an artifact of smoothing a noisy signal. Equivalent spectra \citep{ChasapisApJL17, ChhiberJGR18}, not shown here, have less noise and show this transition at the same scale corresponding to $\sim 24700 km$.}

In MHD, without shears, it is expected that  $\bf v$ and $\bf b$ align increasingly as small scales are approached \citep{BoldyrevPRL06, MasonPRL06, PodestaAIPC08, PodestaJGR09, PodestaApJ10-ch}. The alignment breaks down when kinetic scales are approached \citep{ParasharPRL18, VerdiniApJ18}. Cross helicity changing in the inertial range can be seen in some old Helios observations \citep{TuGRL90,BrunoAIPC96}. \cite{GoldsteinSSPP89, RobertsJGR92} showed that the presence of shears destroys cross helicity at the scales where shear is important. This destruction of measures of Alfv\'enicity deep in the inertial range could potentially be due to large scale or inertial range shear driving that is expected to be important close to the sun.

Even in cases where $\sigma_c$ is fairly constant in the inertial range, other measures could show departures from expected behavior. In Figure \ref{sc_scale08} we show another example of spectra for these measures, in an 8 hour bin centered at 2018-11-02-12:00:00. $\sigma_c$ remains fairly constant in the inertial range down to the noise floor in this particular case, as does $\cos \theta_{vb}$. However, the residual energy and Alfv\'en ratio show a monotonic {reduction in magnetic dominance, as evidenced by the overplotted logarithmic trends. Although in the interval centered at 2018-11-02 12:00:00 flow energy does not dominate for the intervals analyzed, the interval centered at 2018-11-02 4:00:00 transitions into flow energy dominated regime as clearly evidenced by $r_A > 1$.} Hence this interval, although fairly Alfv\'enic at large scales, shows departures from Alfv\'enicity in the sense of energy partition between kinetic and magnetic energies.



Finally, to  ensure that the drop in the inertial range is not affected by noise, we extend the spectral coverage for one of the days by using the data from the flux angle (FA) mode. In this mode, the Faraday cups gather data in a single energy/charge window with 293 Hz cadence. For details of the mode and data processing see \citet{VechApJ19}. The FA mode data are for interval 1 studied in detail in that paper. In Figure \ref{parashar_vech} we show the power spectrum for velocity in the top panel and cross helicity spectra in the bottom panel for the full day of 2018-11-04. The two modes combined cover a spectral range of almost five decades, with FA mode catching up nicely when the noise from SPC becomes significant. The cross helicity at large scales is fairly constant but shows a decline starting about a decade before noise scales are reached. However, the cross helicity computed from the FA mode data nicely continues the logarithmic decay trend for more than a decade below the noise scale for SPC. Combined, the data from these two separate modes of the instrument show a consistent logarithmic decline in cross helicity in most of the inertial range spanned by the two modes. Just before kinetic scales are approached, the cross helicity shows a steep decrease, consistent with what has been observed at 1 au \citep[e.g.][]{ParasharPRL18}. This is indicative of distinct mechanisms responsible for each phase of the decline 
of $\sigma_c$ --  the logarithmic decline in the inertial range, and the steep decline of $\sigma_c$ close to kinetic scales. The former could potentially be because of velocity shears and the latter potentially due to kinetic effects.

\section{Discussion}
Parker Solar Probe (PSP) provides a unique opportunity to study the evolution of heliospheric plasmas close to their place of origin near the Sun. The first perihelion of PSP provides us with a preview into what exciting science lies ahead. Here we have used data from the 
first solar encounter of PSP to study the issue of Alfv\'enicity. The term Alfv\'enic fluctuations carries a wide variety of meanings. In this paper we have studied various possible measures such as cross helicity  $\sigma_c$, residual energy $\sigma_r$, alignment cosine $\cos \theta_{vb}$, and Alfv\'en ratio $r_A$ to quantify the Alfv\'enicity of solar wind near the sun. The fluctuations are Alfv\'enic but not Alfv\'en wave-like. 

Scale decomposition of these quantities is revealing. In some intervals $\sigma_c$ is fairly constant at large scales, indicating the highly Alfv\'enic nature of the interval. However, the scale variations of $\sigma_r$ and $r_A$ show monotonic {reduction in magnetic dominance at large scales, transitioning to flow energy dominated behavior at small scales in one of the intervals.} This indicates a departure from Alfv\'enicity in the energetic sense. In some intervals, even the cross helicity and alignment angles decrease logarithmically deep in the inertial range, unlike what has been observed in the magnetosheath and solar wind at 1 au \citep{ChenApJ13, WicksPRL13, ParasharPRL18} for intervals classically designated as ``Alfv\'enic'' \citep{BelcherJGR71}. {The individual case studies presented here provide motivation for a statistical analysis of Alfvenicity using multiple measures.}

{These case studies suggest} that in such intervals a mechanism other than kinetic physics is acting to reduce the cross-helicity progressively at smaller scales, but still well-removed from kinetic plasma scales. One possibility is the presence of velocity shear driving at large scales that is expected to be significant in the inner heliosphere, and may be present in the outer sub-Alfv\'enic corona. This shear driving could possibly cause a nonlinear Kelvin-Helmholtz-like roll-up at large scales, reducing Alfv\'enicity, and driving a phenomenon that has been described as ``flocculation'' in imaging observations   \citep{DeForestApJ16, ChhiberApJL18}. This possible relation to flocculation will be examined in a separate study. 

\acknowledgments 
The authors acknowledge useful discussions with Chris Chen. This research as been supported 
 in part by the Parker Solar Probe mission under the 
 ISOIS project 
 (contract NNN06AA01C) and a subcontract 
 to University of Delaware from
 Princeton University (SUB0000165).
 Additional support is acknowledged from the  NASA LWS program  (NNX17AB79G), the HSR program 
 (80NSSC18K1210 \& 80NSSC18K1648), and grant RTA6280002 from Thailand Science Research and Innovation. D.V. was supported by NASA's Future Investigators in NASA Earth and Space Science and Technology Program Grant (80NSSC19K1430). Parker Solar Probe was designed, built, and is now operated by the Johns Hopkins Applied Physics Laboratory as part of NASA’s Living with a Star (LWS) program (contract NNN06AA01C). Support from the LWS management and technical team has played a critical role in the success of the Parker Solar Probe mission.

\bibliographystyle{aasjournal} 
\bibliography{alfvenicity}
\end{document}